\documentclass[%
 reprint,
 amsmath,amssymb,
 aps,
prb,
]{revtex4-1}

\usepackage{graphicx}
\usepackage{dcolumn}
\usepackage{bm}

\begin{document}

\preprint{APS/123-QED}

\title{Frequency and power dependence of spin-current emission by spin pumping in a thin film YIG/Pt system
}

\author{V. Castel}
 \email{v.m.castel@rug.nl}
\affiliation{ 
University of Groningen, Physics of nanodevices, Zernike Institute for Advanced Materials, Nijenborgh 4, 9747 AG Groningen, The Netherlands.
}%
\author{J. Ben Youssef}%
\affiliation{ 
Universite de Bretagne Occidentale, Laboratoire de Magnetisme de Bretagne CNRS, 29285 Brest, France.
}%

\author{N. Vlietstra}
\affiliation{ 
University of Groningen, Physics of nanodevices, Zernike Institute for Advanced Materials, Nijenborgh 4, 9747 AG Groningen, The Netherlands.
}%

\author{B. J. van Wees}
\affiliation{ 
University of Groningen, Physics of nanodevices, Zernike Institute for Advanced Materials, Nijenborgh 4, 9747 AG Groningen, The Netherlands.
}%

\date{\today}

\begin{abstract}

This paper presents the frequency dependence of the spin current emission in a hybrid ferrimagnetic insulator/normal metal system. The system is based on a ferrimagnetic insulating thin film of Yttrium Iron Garnet (YIG, 200 nm) grown by liquid-phase-epitaxy (LPE) coupled with a normal metal with a strong spin-orbit coupling (Pt, 15 nm). The YIG layer presents an isotropic behaviour of the magnetization in the plane, a small linewidth, and a roughness lower than 0.4 nm. Here we discuss how the voltage signal from the spin current detector depends on the frequency [0.6 - 7 GHz], the microwave power, $P_{\texttt{in}}$, [1 - 70 mW], and the in-plane static magnetic field. A strong enhancement of the spin current emission is observed at low frequencies, showing the appearance of non-linear phenomena.
\\
\begin{description}

\item[PACS numbers]
72.25.Ba, 72.25.Pn, 75.78.-n, 76.50.+g
\end{description}
\end{abstract}

\keywords{Insulators, Yttrium Iron Garnet (YIG), liquid-phase-epitaxy, Pt, spin pumping, spin hall effect, spin current, broadband ferromagnetic resonance}
\maketitle


\section{Introduction}

The actuation, detection and control of the magnetization dynamics and spin currents in hybrid structures (magnetic material/normal metal) by using the (inverse) spin hall effect (ISHE and SHE), spin transfer torque (STT) and spin pumping, has attracted much attention in the last few years. The observation of these phenomena in ferromagnetic (FM)/normal metal (NM) systems has been reported by several groups \citep{Azevado2005, SaitohAPL2006, AndoPRL2008, AzevedoPRB2011}. 

Spin pumping is the generation of spin currents from magnetization precession, which can be excited by microwave radiation (microstrip\citep{SaitohFreqDep}, resonant cavity\citep{Kajiwara2010nature}, wave-guide\citep{AzevedoYIGthickness}). In a FM/NM system, this spin current is injected into the NM layer, where it is converted into a dc electric voltage using the ISHE. In 2010, Y. Kajiwara et al.\citep{Kajiwara2010nature} opened new interest in this research field by the demonstration of the spin pumping/ISHE and SHE/STT processes in a hybrid system using the magnetic insulating material Yttrium Iron Garnet (YIG, 1.3 $\mu$m), coupled with a thin layer of platinum (Pt, 10 nm). It has been shown experimentally that the combination of these materials and the mentioned phenomena can be used to transmit electrical information over several millimeters \citep{Schneider2008,Kajiwara2010nature,travellingSW}. The insulator/normal metal (YIG/Pt) system presents an important role for future electronic devices related to non-linear dynamics effects\citep{SaitohBistable,Kurebayashi2011nmat, HillebrandsSPmagnons, DemoParametric,2mag}, such as active magnetostatic wave delay lines and signal to noise enhancers, and bistable phenomena\citep{prabhakar4859}.

In this paper, spin current emission in a hybrid structure YIG [200 nm]/Pt [15 nm] as a function of microwave frequency $f$, microwave power $P_{\texttt{in}}$ and applied magnetic field $B$ (in-plane) is presented. The actuation of the spin current emission is provided by a non-resonant 50 $\Omega$ microstrip reflection line\footnote{The characteristic impedance of the microstrip is designed with respect to the source impedance. By taking into account the geometric dimensions of the line, the electrical properties of the line (Au), and the permittivity of the substrate (alumina) we have realized a microstrip line with an impedance of 50 Ohm. In order to create a maximum current through the microstrip and therefore a maximum coupling (in the frequency range that we have used), the end of the transmission line has been shorted. } within a range of $f$ between 0.6 and 7 GHz. To our best knowledge, in all previous experiments, the thickness of the single-crystal of YIG, grown by liquid-phase-epitaxy (LPE), is within a range of 1.3 to 28 $\mu$m, which is always higher than the exchange correlation length defined in pure YIG\citep{Kurebayashi2011nmat}. In contrast, the thickness of the YIG used for the experiments presented here is only 200 nm. Experiments with lower thickness of YIG have been reported\citep{PLD_YIG,WangAPL2011}, however these layers are grown by different methods than LPE. The different growing processes result in an enhancement of the linewidth and these layers do not reach the high quality as when grown by LPE. Besides its thickness, two other points should be made concerning our YIG sample. First the magnetic field (in-plane) dependence of the magnetization presents isotropic behaviour and second, no stripe domains have been observed by Magnetic Force Microscopy (MFM).

\section{Experimental details}

\subsection{Sample description}
Spin pumping experiments in FM/NM systems for different NM materials have been performed in order to study the magnitude of the dc voltage induced by the ISHE \citep{AndoIEEE2010}. It has been shown that the mechanism for spin-charge conversion is effective in metals with strong spin-orbit interaction. Therefore, for the experiments presented in this paper, Pt is used as normal metal layer. As magnetic layer, the insulating material Y$_3$Fe$_5$O$_{12}$ (YIG) is used. The sample is based on a layer of single-crystal Y$_3$Fe$_5$O$_{12}$ (YIG) (111), grown on a (111) Gd$_3$Ga$_5$O$_{12}$ (GGG) single-crystal substrate by liquid-phase-epitaxy (LPE). The thickness of the YIG is only 200 nm, which is very low compared to other studies\citep{Kajiwara2010nature,AzevedoAPL2011,AzevedoPRL2011,Kurebayashi2011nmat, HillebrandsSPmagnons}. The YIG layer has a roughness of 0.4 nm. X-ray diffraction was used in order to estimate the quality of the thin layer of YIG. The spectrum (not shown) shows epitaxial growing of YIG oriented along the (111) direction with zero lattice mismatch. 

\begin{figure}[h]
\includegraphics[width=8cm]{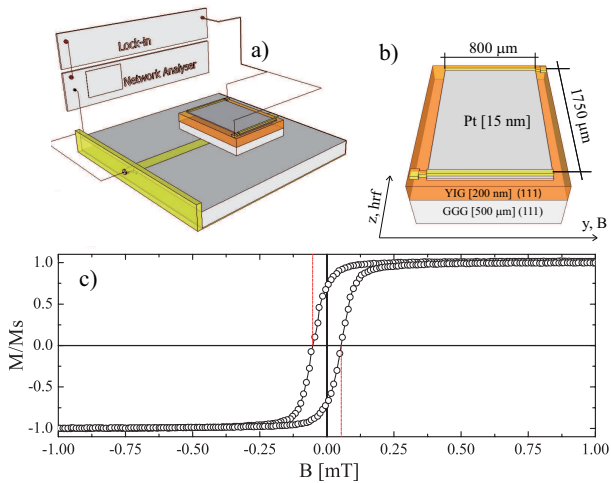}
\caption{\label{fig:Fig1} a) and b) schematics of the experimental setup for spin pumping measurements. The ferromagnetic resonance in the YIG is excited by using a microstrip line in reflection between 0.6 and 7 GHz. The thickness of the YIG and the GGG substrate is 200 nm and 500 $\mu$m, respectively. Ti/Au electrodes are attached on top of the Pt layer in order to detect the ISHE voltage. The magnetic field $B$ is applied in the plane of the sample along the $y$ direction and $B\perp h_{\texttt{rf}}$, where $h_{\texttt{rf}}$ is the microwave field. c) magnetic field (in-plane) dependence of the magnetization $M$ (normalized by $M_{s}$, the saturation magnetization) of the pure single-crystal of YIG performed by Vibrating Sample Magnetometer (VSM) at room temperature.}
\end{figure}

For the realization of the hybrid structure, two steps of lithography have been used. First, to create the Pt layer (15 nm thick), an area of 800$\times$1750 $\mu$m$^{2}$ has been patterned on top of a YIG sample (1500$\times$3000 $\mu$m$^{2}$), by electron beam lithography (EBL). Before deposition of the Pt layer by dc sputtering, argon etching has been used to clean the surface. Etching was done during 5 seconds at a beam voltage (intensity) of 500 V (14 mA) with an acceleration voltage of 200 V. The second lithography step realizes the Ti/Au electrodes of 30 $\mu$m width and 100 nm thick. For both lithography steps, PMMA with a thickness of 270 nm has been used as resist. A schematic of the final device is shown in Fig.\ref{fig:Fig1} b).

\subsection{Static and dynamic magnetization characterizations}
By using specific growing conditions, the anisotropic contributions (growth, and magneto-elastic) in the YIG film can be optimized in order to keep the magnetization in-plane. Fig.\ref{fig:Fig1} c) shows the dependence of the longitudinal component of the magnetization as a function of the magnetic field applied in the plane of the YIG sample, as measured by using a Vibrating Sample Magnetometer (VSM) at room temperature. The saturation magnetization is $\mu_{0}M$ =0.176 T, corresponding to the value obtained for YIG in bulk \citep{Kajiwara2010nature,Kurebayashi2011nmat}. The low coercive field ($\simeq$ 0.06 mT) and the shape of the hysteresis loop provide an easy proof of the magnetization being in the plane, with a very low dissipation of the energy. VSM measurements along the two crystallographic axis, [1,$\bar{1}$,0] and [1,1,$\bar{2}$], show similar responses indicating isotropic behaviour of the magnetization in the film plane. In addition, no stripe domains have been observed by MFM. 

In order to well characterize the pure single-crystal of YIG, before realizing the YIG/Pt structure, broadband ferromagnetic resonance (FMR) measurements have been performed using a highly sensitive wideband resonance spectrometer in the perpendicular configuration (the applied magnetic field, $B$, is normal to the film plane). The microwave excitation is provided with a non-resonant 50 $\Omega$ microstrip reflection line within a range of microwave frequencies between 2 and 25 GHz. The FMR is measured via the first derivative of the power absorption $dP/dH$ by using a lock-in measurement technique. The value of the modulation field (lock-in reference) used during the field sweeping is much smaller than the FMR linewidth. The dependence of the frequency resonance, $\omega_{\texttt{res}}$, as a function of the resonant magnetic field is used to determine the gyromagnetic ratio $\gamma$=1.80 10$^{11}$ radT$^{-1}$s$^{-1}$ (and the Lande factor, $g$=2.046). The intrinsic Gilbert damping parameter is extracted from the dependence of the linewidth as a function of the microwave frequency ($\alpha\approx$ 2 10$^{-4}$)\citep{jbyJAP2010}.

\subsection{Spin pumping measurement}
For the actuation of the magnetization resonance in the YIG layer, a different FMR setup has been used. To connect the device, the YIG/Pt system is placed as shown in Fig.\ref{fig:Fig1} a). In this configuration, the microwave field $h_{\texttt{rf}}$ is perpendicular to the static magnetic field, $B$. To optimize the electric voltage recording, a lock-in measurement technique was used. The frequency reference, generated by the lock-in, is send to the network analyser trigger. This command (with a frequency of 17 Hz) controls the microwave field by the network analyser. The microwave field is periodically switched on and off between $P^{ \texttt{High}}_{\texttt{rf}}$ and $P^{\texttt{Low}}_{\texttt{rf}}$, respectively. $P^{\texttt{Low}}_{\texttt{rf}}$ is equal to 0.001 mW and $P^{ \texttt{High}}_{\texttt{rf}}$ corresponds to the input microwave power, so-called in the following, $P_{\texttt{in}}$. The dc voltages generated between the edges of the Pt layer are amplified and detected as a difference of $V(P^{ \texttt{High}}_{\texttt{rf}})-V(P^{\texttt{Low}}_{\texttt{rf}})$. 

Using this measurement setup, the dependence of the electric voltage signal as a function of the microwave power [1-70 mW] and the frequency [0.6-7 GHz] is analysed, while sweeping the applied static magnetic field, $B$. $B$ is large enough in order to saturate the magnetization along the plane film. All measurements were performed at room temperature.

\section{Results and discussion}
Conversion of spin currents into electric voltage via the ISHE is given by the relation \citep{Kajiwara2010nature}: $E_{\texttt{ISHE}}\propto J_{s}\times \sigma$, where $E_{\texttt{ISHE}}$, and $ \sigma $ are the electric field induced by the ISHE and the spin polarization, respectively.
\begin{figure}[h]
\includegraphics[width=9cm]{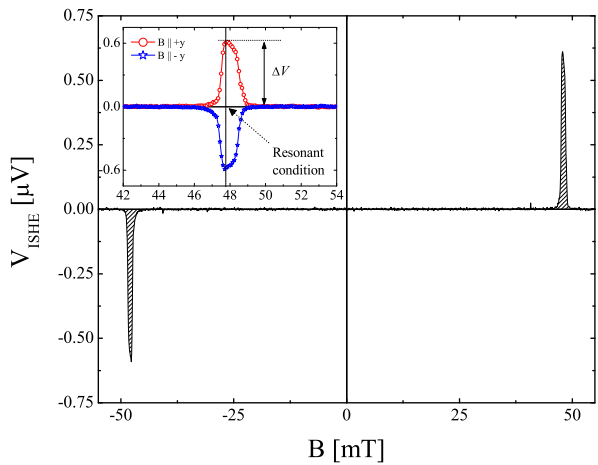}
\caption{\label{fig:Fig2} Dependence of the electric voltage signal, $V_{\texttt{ISHE}}$, as a function of the magnetic field, $B$, for the YIG [200 nm]/Pt [15 nm] sample. $B$ is applied in-plane and the microwave parameters are fixed at 3 GHz and 20 mW. The inset shows $V_{\texttt{ISHE}}$ at resonant condition for the positive and negative configuration of the magnetic field (along $+y$ and $-y$, respectively, see Fig.\ref{fig:Fig1}b)).}
\end{figure}
In YIG/Pt, the origin of the spin current, $ J_{\texttt{s}} $, injected through the Pt layer differs from the conventional spin current in conducting systems like Py/Pt. The spin pumping originates from the spin exchange interaction between a localized moment in YIG at the interface and a conduction electron in the Pt layer.

The magnetic field dependence of the voltage signal in YIG [200 nm]/Pt [15nm] at 3 GHz is shown in Fig. \ref{fig:Fig2}. The rf microwave power is fixed at 20 mW. The sign of the electric voltage signal is changed \citep{Kajiwara2010nature} by reversing the magnetic field along $ y $ and no sizeable voltage is measured when $B$ is parallel to $z$, as expected. The reversing of the sign of $V$ (by reversing the magnetic field) shows that the measured signal is not produced by a possible thermoelectric effect, induced by the microwave absorption. A direct measurement of the electric voltage signal (without lock-in amplifier) has been performed in order to define the sign of $V_{\texttt{ISHE}}$ as a function of the magnetic and electric configuration. The voltage detected between the edges of the Pt layer shows resonance-like behaviour, with a maximum value ($ \Delta V $) at the resonant condition of the system as defined in the inset of Fig.\ref{fig:Fig2}. 

\begin{figure}[h]
\includegraphics[width=8cm]{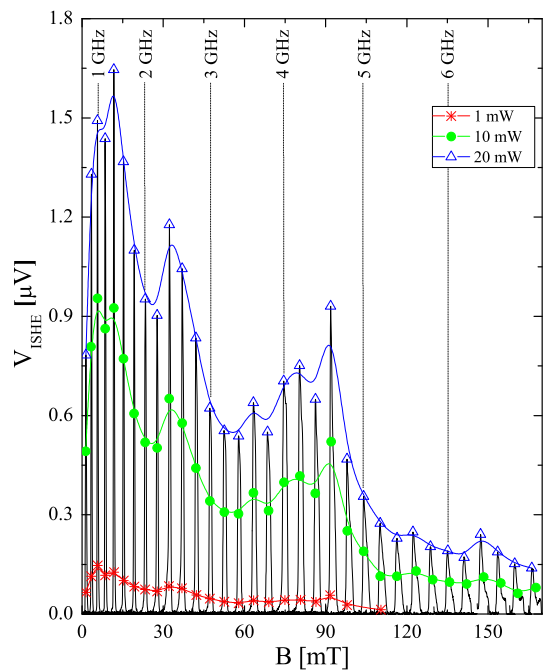}
\caption{\label{fig:Fig3} Dependence of the electric voltage signal, $V_{\texttt{ISHE}}$, for the YIG [200 nm]/Pt [15 nm] sample as a function of the static magnetic field (in-plane) within a microwave frequency range of [0.6-7 GHz] at 20 mW. Symbols correspond to the value of $\Delta V$ for different microwave power: 1, 10, and 20 mW.}
\end{figure}

In Fig.\ref{fig:Fig3} the in-plane magnetic field dependence of the electric voltage signal for a large range of microwave frequencies between 0.6 and 7 GHz is shown. For each value of microwave power ($P_{\texttt{in}}$=1, 10, and 20 mW) and frequency, $f$, the voltage signal, $V_{\texttt{ISHE}}$=f($P_{\texttt{in}}$, $f$), at resonant conditions has been extracted. To our best knowledge, only two groups\citep{Kajiwara2010nature,Kurebayashi2011nmat} have studied the electric voltage signal in a hybrid YIG/Pt system as a function of microwave frequency, but only one\citep{Kurebayashi2011nmat} in a large frequency range of [2-6.8 GHz]. The difference between our structure and Ref.\citep{Kurebayashi2011nmat} lies in the thickness of the YIG, which is 5.1 $\mu$m in their case and only 200 nm in this work. The thickness of the Pt is the same (15 nm). 
As can be seen from Fig.\ref{fig:Fig3}, the frequency dependence of $\Delta V$ presents a complicated evolution, partly resulting from the $S_{11}$ dependence of the microstrip in reflection itself, as a function of frequency. Nevertheless, note that $\Delta V$ presents high values at low frequency.

\begin{figure}
\includegraphics[width=9cm]{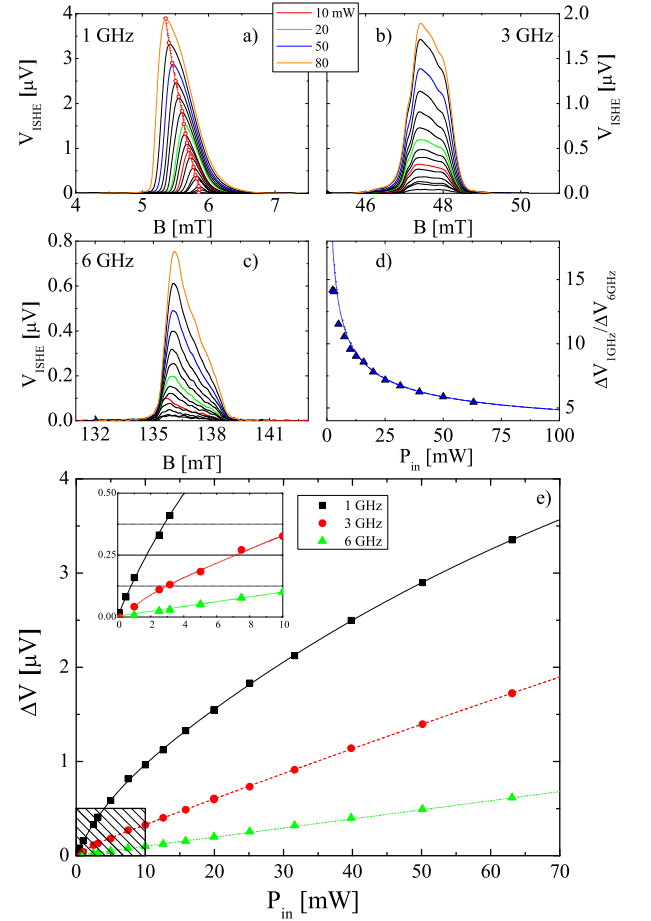}
\caption{\label{fig:Fig4}a), b), and c) present the dependence of the electric voltage signal, $V_{\texttt{ISHE}}$, as a function of the static magnetic field, $B$, for different microwave power, $P_{\texttt{in}}$, at 1, 3, and 6 GHz, respectively. d) microwave power dependence of the ratio of the values of $\Delta V$ measured at 1 and 6 GHz. e) representation of $\Delta V$ as a function of the microwave power between 1 and 70 mW at 1, 3, and 6 GHz. The inset corresponds to the dependence of $\Delta V$ for low rf power.}
\end{figure}

Fig.\ref{fig:Fig4} a), b), and c), present the dependence of the electric voltage $V_{\texttt{ISHE}}$ as a function of the magnetic field and the microwave power for different frequencies (1, 3, and 6 GHz, respectively). Two points should be made regarding these graphs. First, one can see in those spectra multiple resonance signals, which are attributed to the Magnetostatic Surface Spin waves (MSSW, when the magnetic field is lower than the resonant condition) and Backward Volume Magnetostatic Spin Waves (BVMSW, when the magnetic field is higher than the resonant condition)\citep{SWstancil,HilleSelection}. Second, the strong non-linear dependence observed at low frequency is well represented by the resonance magnetic field shift and the asymmetric distortion of the resonance line as observed in Fig.\ref{fig:Fig4} a). These observations are correlated with the pioneering works of Suhl\citep{Suhl} and Weiss\citep{Weiss} related to non-linear phenomena occurring at large precession angles. The simple expression\citep{thetaAngle} of the magnetization precession cone angle at resonance is given by $\Theta=h_{rf}/\Delta H$, where $h_{rf}$ and $\Delta H$ correspond to the microwave magnetic field and the linewidth of the absorption line of the uniform mode, respectively. This expression shows that by decreasing the excitation frequency, an enhancement of the cone angle is induced. Therefore, the system becomes more sensitive to the rf microwave power, $P_{\texttt{in}}$.

In addition, the non-linear behaviour measured at 1 GHz (also at 3 GHz, but less) is well represented by Fig.\ref{fig:Fig4} e). This figure represents evolutions of $\Delta V$ as a function of the microwave power, $P_{\texttt{in}}$, performed at 1, 3 and 6 GHz between 1 and 70 mW. Y. Kajiwara et al. \citep{Kajiwara2010nature} have proposed an equation to represent the dependence of the electric voltage signal as a function of $B$, $f$, $h_{\texttt{ac}}$ (microwave magnetic field), and the parameters of the bilayer system. They showed that $V_{\texttt{ISHE}}$ at resonant conditions depends linearly on the microwave power. This dependence is well reproduced only at 6 GHz. Fig.\ref{fig:Fig4} d) represents the ratio of $\Delta V$ extracted from measurements at 1 and 6 GHz, $\Delta V_{\texttt{1GHz}}/\Delta V_{\texttt{6GHz}}$, as a function of the microwave power $P_{\texttt{in}}$, to emphasize the non-linearity observed at 1 GHz. Note that, for a very low microwave power of 1 mW, $\Delta V$ at 1 GHz is 14 times greater than $\Delta V$ at 6 GHz, whereas by increasing $P_{\texttt{in}}$, this difference is drastically reduced\citep{Kurebayashi2011nmat} and reached a factor of 5 at 60 mW.

To investigate the frequency dependence of $\Delta V$, the response of the microstrip line should be taken into account. Between 30 and 40 mT and between 70 and 100 mT, the microstrip line induces an artificial increase of $V_{\texttt{ISHE}}$, as can be observed in Fig.\ref{fig:Fig3}. The correction factor for this artificial increase is determined by measuring the reflection parameter $ S_{11}$, for the system being out of resonance. 

\begin{figure} [h]
\includegraphics[width=8cm]{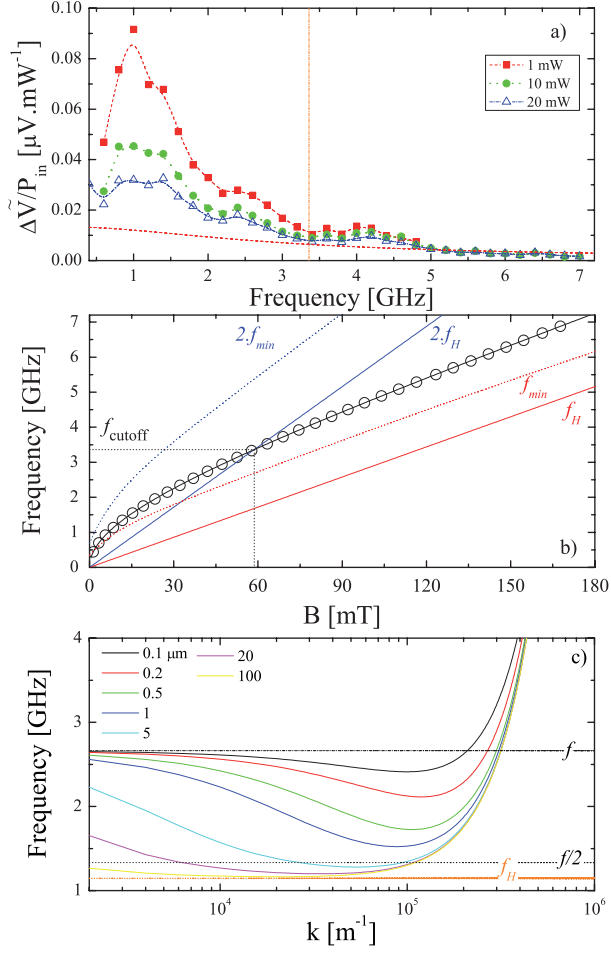}
\caption{\label{fig:Fig5} a) dependence of $\Delta\tilde{V}/P_{\texttt{in}}$ as a function of the microwave frequency with a microwave power of 1, 10, and 20 mW. The red dashed line corresponds to the analytical expression of the frequency dependence of $\Delta V$ extracted from Ref.\citep{Kajiwara2010nature}. b) dependence of the resonant frequency, $f$, as a function of the applied magnetic field. Open circles indicate the experimental data when $k\perp B$ (in-plane magnetic field) and the solid black curve is calculated from the Kittel's formula \citep{Kittel} given by: $f=\sqrt{f_{H}(f_{H}+f_{M})}$. Note that $f_{H}=\gamma\mu_{0} H$ and $f_{M}=\gamma\mu_{0} M$. c) dispersion relation of spin waves\citep{SWdispExchange}: dependence of the frequency as function of the wavevector, $k$, when $k\parallel B$ for different thickness of YIG. The magnetic field is fixed at 40 mT.}
\end{figure}

Fig.\ref{fig:Fig5} a) represents the frequency dependence of $\Delta\tilde{V}/P_{\texttt{in}}$, where $\Delta\tilde{V}$ corresponds to the the dc voltage corrected by the response of the microstrip line itself. This figure permits to define the frequency range in which this evolution presents non-linear behaviour. Note that between 3.4 and 7 GHz, values of $\Delta\tilde{V}/P_{\texttt{in}}$ present a slow decrease as a function of the microwave frequency. In this regime, $\Delta\tilde{V}/P_{\texttt{in}}$ values are similar for the different rf microwave powers of 1, 10, and 20 mW due to the fact that in this frequency range, the rf power dependence of $\Delta V$ is linear\citep{Kajiwara2010nature,DemoParametric}.
The interesting feature of the frequency dependence of $\Delta\tilde{V}/P_{\texttt{in}}$ is observed at frequencies below 3.4 GHz. At those frequencies, the frequency dependence does not follow the trend observed at higher frequencies ($>$3.4 GHz). In the frequency range [0.6-3.4 GHz], the previously observed non-linear behaviour affects the values of $\Delta\tilde{V}/P_{\texttt{in}}$ as a function of the input rf microwave power. The enhancement $\Delta\tilde{V}/P_{\texttt{in}}$ is more efficient at low powers and gradually reduces with increasing the microwave power. The discrepancy is especially strong around the maximum at 1 GHz. H. Kurebayashi et al. \citep{Kurebayashi2011nmat} obtained 17.1 and 62.8 nV/mm at 2 and 6 GHz, respectively, whereas in our system, for the same frequencies, $\Delta V$ reaches 542.85 and 108.6 nV/mm. 

The question arising now: what is the origin of the strong enhancement of $\Delta V$ at low frequency. Is it only due to the frequency dependence of the cone angle? The assumption of a single magnetization precession angle is not warranted, due to the fact that several spin wave modes contribute to the dynamic response of the system. Therefore, assuming that no spin waves are created in the YIG, the normalization of $\Delta V$ by $\Theta$ (defined by $\alpha$) and $P$ cannot explain the enhancement of $\Delta\tilde{V}/P_{\texttt{in}}$ at low frequency. Here, $P$ corresponds to the correction factor related to the ellipticity trajectory of the magnetization precession of the uniform mode\citep{andoOptimum} due to the magnetic field configuration (in-plane). The analytical expression (red dash line in Fig.\ref{fig:Fig5}a)) extracted from Ref.\citep{Kajiwara2010nature}, in which the spin current at the YIG/Pt interface is defined by the uniform mode, cannot reproduce the dc voltage behaviour at low frequency. As reported previously in Ref.\citep{HillebrandsSPmagnons,Kurebayashi2011nmat}, this behaviour has been attributed to the presence of non-linear phenomena. H. Kurebayashi et al.\citep{Kurebayashi2011nmat} have demonstrated the possibility to control the spin current at the YIG/Pt interface by three-magnon splitting. This non-linear phenomenon can be easily actuated for very low rf power\citep{PowerThresNikitov1988}. Kurebayashi et al.\citep{Kurebayashi2011nmat} have observed that the threshold power of the splitting in their system was around 18 $\mu$W, which is very low with respect to the rf power used for FMR and dc voltage measurements.

Fig.\ref{fig:Fig5} b) introduces the frequency limit of the three-magnon splitting boundaries calculated for our sample (200 nm) and for a thick sample of YIG. The splitting induces the creation of two magnons (with short-wavelength) from the uniform mode (long-wavelength), following the equations: $f=f_{1}+f_{2}$ and $k=k_{1}+k_{2}$, where $f $ and $k$ are the frequency and wave vector with $f_{1}=f_{2}=\frac{1}{2}f$ and $k_{1}=-k_{2}$\citep{SWstancil, PowerThresNikitov1988}. In agreement with Kurebayashi et al.\citep{Kurebayashi2011nmat}, a strong enhancement of the dc voltage at low frequency has been observed, but this dependence does not necessarily mean that three-magnon splitting is involved in our system. By following the schema of the three-magnon, one can easily see that this phenomena is allowed for a specific frequency range. The upper frequency limit ($f_{\texttt{cutoff}}$) for the splitting is defined by the minimum of the BVSWM dispersion curve ($f_{\texttt{min}}$) resulting from the competition between the dipole interaction and the exchange interaction. This minimum depends of the thickness of the YIG sample. 

For a thick sample of YIG, $f_{\texttt{min}}\approx f_{H}$, where $f_{H}$ is the Larmor frequency. The FMR frequency cannot be lower than $f_{H}$ \citep{TMSNikitov}, and thus, the excitation frequency should be higher than $2f_{H}$ in order that the process described by the above equation can take place. Consequently, the upper frequency limit, $f_{\texttt{cutoff}}$, for a thick sample system of YIG is $f_{\texttt{cutoff}}=\dfrac{2}{3}f_{M}$, where $f_{M}=\gamma\mu_{0} M$. In the experiment of Kurebayashi et al.\citep{Kurebayashi2011nmat}, they have used a YIG sample with a thickness of 5.1 $\mu$m, which is higher than the exchange correlation length, and therefore $f_{\texttt{min}}\approx f_{H}$. 

Nevertheless, by taking into account a YIG thickness of 200 nm, the dependence of $f_{\texttt{min}}$ (from Ref.\citep{SWdispExchange}) shows a strong difference with $f_{H}$ (see Fig.\ref{fig:Fig5} b)). The model of the three-magnon splitting ($f>2f_{\texttt{min}}$) suggested that in our case this process is not allowed. Fig.\ref{fig:Fig5} c) represents the spin wave spectrum in YIG when the magnetic field is parallel to the wavevector, $k$, for different thickness of YIG\citep{SWdispExchange}. The calculation has been performed with a magnetic field of 40 mT inducing a microwave frequency $f$=2.66 GHz with $\gamma$=1.80 10$^{11}$ rad T$^{-1}$s$^{-1}$. A crossing of the dispersion curve with the black dotted line ($f/2$) shows that the splitting is permitted. By reducing the thickness, the minimum frequency increases. For thin layers of YIG, the dispersion curve does not cross the black dotted line anymore, suggesting that here the three-magnon splitting is no longer allowed. 

The role of the three-magnon splitting process for the spin pumping is not fully clear and there are many non-linear phenomena which can induce the creation of spin waves with short-wavelength (multi-magnon processes such as four-magnon and two-magnon scattering). It has been shown by Jungfleisch et al.\citep{2mag} that the two-magnon process (due to the scattering of magnons on impurities and surfaces of the film) contributes to enhance the spin current at the YIG/Pt interface. The strong enhancement of $\Delta V$ observed at low frequency is due to the fact that the dc voltage induced by spin pumping at the YIG/Pt interface is insensitive to the spin waves wavelength\citep{Kurebayashi2011nmat,2mag}. In other words, $\Delta V$ is not only defined by the uniform mode but from secondary spin wave modes, which present short-wavelength. It is not obvious to identify the contributions of the different multi-magnon processes, involved in our system, to the enhancement of the dc voltage at low frequency.

\section{Conclusion}

In summary, we have shown spin current emission in a hybrid structure YIG [200 nm]/Pt [15 nm] as a function of microwave frequency $f$, microwave power $P_{\texttt{in}}$ and applied magnetic field $B$ (in-plane). We have observed a strong enhancement of the voltage signal emission across a spin current detector of Pt at low frequency. This behaviour can be understood if we assume that the measured signal is not only driven by the FMR mode (which contributes to the spin-pumping at the YIG/Pt interface) but also from a spectrum of secondary spin-wave modes, presenting short wavelengths. 

In YIG-based electronic devices, the creation of short-wavelength spin waves is considered as a parasitic effect. However, in this case it can be used as a spin current amplifier. Before to integrate this system in a device, many questions related to the contribution for the spin pumping of the spin waves with short-wavelength should be solved. To date, no systematic studies of the spin current emission on a YIG/Pt system have been done as function of the YIG thickness. By choosing a specific thickness range, it should be possible to follow the contribution of the three-magnon splitting ($f_{\texttt{cutoff}}$) by a combination of Brillouin Light Scattering (BLS)\citep{Kurebayashi2011nmat} and spin pumping measurements. More details of other multi-magnon processes should be given by temperature dependence measurements. Nevertheless, the enhancement of $\Delta V$, which we have observed in the frequency range [0.6 - 3.2 GHz], could be used to downscale a hybrid structure of YIG/Pt. The isotropic behaviour of the in-plane magnetization, the absence of stripe domains, and the high quality thin layer of YIG (200 nm) grown by liquid-phase-epitaxy give keys to success in this way. 

We would like to acknowledge N. Vukadinovic for valuable discussions and B. Wolfs, M. de Roosz and J. G. Holstein for technical assistance. This work is part of the research program (Magnetic Insulator Spintronics) of the Foundation for Fundamental Research on Matter (FOM) and supported by NanoNextNL, a micro and nanotechnology consortium of the Government of the Netherlands and 130 partners (06A.06), NanoLab NL and the Zernike Institute for Advanced Materials.


%


\bibliography{YIGPt}

\providecommand{\noopsort}[1]{}\providecommand{\singleletter}[1]{#1}%
\begin{thebibliography}{32}%
\makeatletter
\providecommand \@ifxundefined [1]{%
 \@ifx{#1\undefined}
}%
\providecommand \@ifnum [1]{%
 \ifnum #1\expandafter \@firstoftwo
 \else \expandafter \@secondoftwo
 \fi
}%
\providecommand \@ifx [1]{%
 \ifx #1\expandafter \@firstoftwo
 \else \expandafter \@secondoftwo
 \fi
}%
\providecommand \natexlab [1]{#1}%
\providecommand \enquote  [1]{``#1''}%
\providecommand \bibnamefont  [1]{#1}%
\providecommand \bibfnamefont [1]{#1}%
\providecommand \citenamefont [1]{#1}%
\providecommand \href@noop [0]{\@secondoftwo}%
\providecommand \href [0]{\begingroup \@sanitize@url \@href}%
\providecommand \@href[1]{\@@startlink{#1}\@@href}%
\providecommand \@@href[1]{\endgroup#1\@@endlink}%
\providecommand \@sanitize@url [0]{\catcode `\\12\catcode `\$12\catcode
  `\&12\catcode `\#12\catcode `\^12\catcode `\_12\catcode `\%12\relax}%
\providecommand \@@startlink[1]{}%
\providecommand \@@endlink[0]{}%
\providecommand \url  [0]{\begingroup\@sanitize@url \@url }%
\providecommand \@url [1]{\endgroup\@href {#1}{\urlprefix }}%
\providecommand \urlprefix  [0]{URL }%
\providecommand \Eprint [0]{\href }%
\providecommand \doibase [0]{http://dx.doi.org/}%
\providecommand \selectlanguage [0]{\@gobble}%
\providecommand \bibinfo  [0]{\@secondoftwo}%
\providecommand \bibfield  [0]{\@secondoftwo}%
\providecommand \translation [1]{[#1]}%
\providecommand \BibitemOpen [0]{}%
\providecommand \bibitemStop [0]{}%
\providecommand \bibitemNoStop [0]{.\EOS\space}%
\providecommand \EOS [0]{\spacefactor3000\relax}%
\providecommand \BibitemShut  [1]{\csname bibitem#1\endcsname}%
\let\auto@bib@innerbib\@empty
\bibitem [{\citenamefont {Azevedo}\ \emph {et~al.}(2005)\citenamefont
  {Azevedo}, \citenamefont {Vilela-Le\~ao}, \citenamefont
  {Rodr\'iguez-Su\'arez}, \citenamefont {Oliveira},\ and\ \citenamefont
  {Rezende}}]{Azevado2005}%
  \BibitemOpen
  \bibfield  {author} {\bibinfo {author} {\bibfnamefont {A.}~\bibnamefont
  {Azevedo}}, \bibinfo {author} {\bibfnamefont {L.~H.}\ \bibnamefont
  {Vilela-Le\~ao}}, \bibinfo {author} {\bibfnamefont {R.~L.}\ \bibnamefont
  {Rodr\'iguez-Su\'arez}}, \bibinfo {author} {\bibfnamefont {A.~B.}\
  \bibnamefont {Oliveira}}, \ and\ \bibinfo {author} {\bibfnamefont {S.~M.}\
  \bibnamefont {Rezende}},\ }\href {\doibase 10.1063/1.1855251} {\bibfield
  {journal} {\bibinfo  {journal} {Journal of Applied Physics}\ }\textbf
  {\bibinfo {volume} {97}},\ \bibinfo {eid} {10C715} (\bibinfo {year}
  {2005})}\BibitemShut {NoStop}%
\bibitem [{\citenamefont {Saitoh}\ \emph {et~al.}(2006)\citenamefont {Saitoh},
  \citenamefont {Ueda}, \citenamefont {Miyajima},\ and\ \citenamefont
  {Tatara}}]{SaitohAPL2006}%
  \BibitemOpen
  \bibfield  {author} {\bibinfo {author} {\bibfnamefont {E.}~\bibnamefont
  {Saitoh}}, \bibinfo {author} {\bibfnamefont {M.}~\bibnamefont {Ueda}},
  \bibinfo {author} {\bibfnamefont {H.}~\bibnamefont {Miyajima}}, \ and\
  \bibinfo {author} {\bibfnamefont {G.}~\bibnamefont {Tatara}},\ }\href
  {\doibase 10.1063/1.2199473} {\bibfield  {journal} {\bibinfo  {journal}
  {Applied Physics Letters}\ }\textbf {\bibinfo {volume} {88}},\ \bibinfo {eid}
  {182509} (\bibinfo {year} {2006})}\BibitemShut {NoStop}%
\bibitem [{\citenamefont {Ando}\ \emph {et~al.}(2008)\citenamefont {Ando},
  \citenamefont {Takahashi}, \citenamefont {Harii}, \citenamefont {Sasage},
  \citenamefont {Ieda}, \citenamefont {Maekawa},\ and\ \citenamefont
  {Saitoh}}]{AndoPRL2008}%
  \BibitemOpen
  \bibfield  {author} {\bibinfo {author} {\bibfnamefont {K.}~\bibnamefont
  {Ando}}, \bibinfo {author} {\bibfnamefont {S.}~\bibnamefont {Takahashi}},
  \bibinfo {author} {\bibfnamefont {K.}~\bibnamefont {Harii}}, \bibinfo
  {author} {\bibfnamefont {K.}~\bibnamefont {Sasage}}, \bibinfo {author}
  {\bibfnamefont {J.}~\bibnamefont {Ieda}}, \bibinfo {author} {\bibfnamefont
  {S.}~\bibnamefont {Maekawa}}, \ and\ \bibinfo {author} {\bibfnamefont
  {E.}~\bibnamefont {Saitoh}},\ }\href {\doibase
  10.1103/PhysRevLett.101.036601} {\bibfield  {journal} {\bibinfo  {journal}
  {Phys. Rev. Lett.}\ }\textbf {\bibinfo {volume} {101}},\ \bibinfo {pages}
  {036601} (\bibinfo {year} {2008})}\BibitemShut {NoStop}%
\bibitem [{\citenamefont {Azevedo}\ \emph {et~al.}(2011)\citenamefont
  {Azevedo}, \citenamefont {Vilela-Le\~ao}, \citenamefont
  {Rodr\'iguez-Su\'arez}, \citenamefont {Lacerda~Santos},\ and\ \citenamefont
  {Rezende}}]{AzevedoPRB2011}%
  \BibitemOpen
  \bibfield  {author} {\bibinfo {author} {\bibfnamefont {A.}~\bibnamefont
  {Azevedo}}, \bibinfo {author} {\bibfnamefont {L.~H.}\ \bibnamefont
  {Vilela-Le\~ao}}, \bibinfo {author} {\bibfnamefont {R.~L.}\ \bibnamefont
  {Rodr\'iguez-Su\'arez}}, \bibinfo {author} {\bibfnamefont {A.~F.}\
  \bibnamefont {Lacerda~Santos}}, \ and\ \bibinfo {author} {\bibfnamefont
  {S.~M.}\ \bibnamefont {Rezende}},\ }\href {\doibase
  10.1103/PhysRevB.83.144402} {\bibfield  {journal} {\bibinfo  {journal} {Phys.
  Rev. B}\ }\textbf {\bibinfo {volume} {83}},\ \bibinfo {pages} {144402}
  (\bibinfo {year} {2011})}\BibitemShut {NoStop}%
\bibitem [{\citenamefont {Harii}\ \emph {et~al.}(2011)\citenamefont {Harii},
  \citenamefont {An}, \citenamefont {Kajiwara}, \citenamefont {Ando},
  \citenamefont {Nakayama}, \citenamefont {Yoshino},\ and\ \citenamefont
  {Saitoh}}]{SaitohFreqDep}%
  \BibitemOpen
  \bibfield  {author} {\bibinfo {author} {\bibfnamefont {K.}~\bibnamefont
  {Harii}}, \bibinfo {author} {\bibfnamefont {T.}~\bibnamefont {An}}, \bibinfo
  {author} {\bibfnamefont {Y.}~\bibnamefont {Kajiwara}}, \bibinfo {author}
  {\bibfnamefont {K.}~\bibnamefont {Ando}}, \bibinfo {author} {\bibfnamefont
  {H.}~\bibnamefont {Nakayama}}, \bibinfo {author} {\bibfnamefont
  {T.}~\bibnamefont {Yoshino}}, \ and\ \bibinfo {author} {\bibfnamefont
  {E.}~\bibnamefont {Saitoh}},\ }\href {\doibase 10.1063/1.3594661} {\bibfield
  {journal} {\bibinfo  {journal} {Journal of Applied Physics}\ }\textbf
  {\bibinfo {volume} {109}},\ \bibinfo {eid} {116105} (\bibinfo {year}
  {2011})}\BibitemShut {NoStop}%
\bibitem [{\citenamefont {Kajiwara}\ \emph {et~al.}(2010)\citenamefont
  {Kajiwara}, \citenamefont {Harii}, \citenamefont {Takahashi}, \citenamefont
  {Ohe}, \citenamefont {Uchida}, \citenamefont {Mizuguchi}, \citenamefont
  {Umezawa}, \citenamefont {Kawai}, \citenamefont {Ando}, \citenamefont
  {Takanashi}, \citenamefont {Maekawa},\ and\ \citenamefont
  {Saitoh}}]{Kajiwara2010nature}%
  \BibitemOpen
  \bibfield  {author} {\bibinfo {author} {\bibfnamefont {Y.}~\bibnamefont
  {Kajiwara}}, \bibinfo {author} {\bibfnamefont {K.}~\bibnamefont {Harii}},
  \bibinfo {author} {\bibfnamefont {S.}~\bibnamefont {Takahashi}}, \bibinfo
  {author} {\bibfnamefont {J.}~\bibnamefont {Ohe}}, \bibinfo {author}
  {\bibfnamefont {K.}~\bibnamefont {Uchida}}, \bibinfo {author} {\bibfnamefont
  {M.}~\bibnamefont {Mizuguchi}}, \bibinfo {author} {\bibfnamefont
  {H.}~\bibnamefont {Umezawa}}, \bibinfo {author} {\bibfnamefont
  {H.}~\bibnamefont {Kawai}}, \bibinfo {author} {\bibfnamefont
  {K.}~\bibnamefont {Ando}}, \bibinfo {author} {\bibfnamefont {K.}~\bibnamefont
  {Takanashi}}, \bibinfo {author} {\bibfnamefont {S.}~\bibnamefont {Maekawa}},
  \ and\ \bibinfo {author} {\bibfnamefont {E.}~\bibnamefont {Saitoh}},\ }\href
  {\doibase 10.1038/nature08876} {\bibfield  {journal} {\bibinfo  {journal}
  {Nature (London)}\ }\textbf {\bibinfo {volume} {464}},\ \bibinfo {pages}
  {262} (\bibinfo {year} {2010})}\BibitemShut {NoStop}%
\bibitem [{\citenamefont {Vilela-Le\~ao}\ \emph {et~al.}(2011)\citenamefont
  {Vilela-Le\~ao}, \citenamefont {Salvador}, \citenamefont {Azevedo},\ and\
  \citenamefont {Rezende}}]{AzevedoYIGthickness}%
  \BibitemOpen
  \bibfield  {author} {\bibinfo {author} {\bibfnamefont {L.~H.}\ \bibnamefont
  {Vilela-Le\~ao}}, \bibinfo {author} {\bibfnamefont {C.}~\bibnamefont
  {Salvador}}, \bibinfo {author} {\bibfnamefont {A.}~\bibnamefont {Azevedo}}, \
  and\ \bibinfo {author} {\bibfnamefont {S.~M.}\ \bibnamefont {Rezende}},\
  }\href {\doibase 10.1063/1.3631683} {\bibfield  {journal} {\bibinfo
  {journal} {Applied Physics Letters}\ }\textbf {\bibinfo {volume} {99}},\
  \bibinfo {eid} {102505} (\bibinfo {year} {2011})}\BibitemShut {NoStop}%
\bibitem [{\citenamefont {Schneider}\ \emph {et~al.}(2008)\citenamefont
  {Schneider}, \citenamefont {Serga}, \citenamefont {Leven}, \citenamefont
  {Hillebrands}, \citenamefont {Stamps},\ and\ \citenamefont
  {Kostylev}}]{Schneider2008}%
  \BibitemOpen
  \bibfield  {author} {\bibinfo {author} {\bibfnamefont {T.}~\bibnamefont
  {Schneider}}, \bibinfo {author} {\bibfnamefont {A.~A.}\ \bibnamefont
  {Serga}}, \bibinfo {author} {\bibfnamefont {B.}~\bibnamefont {Leven}},
  \bibinfo {author} {\bibfnamefont {B.}~\bibnamefont {Hillebrands}}, \bibinfo
  {author} {\bibfnamefont {R.~L.}\ \bibnamefont {Stamps}}, \ and\ \bibinfo
  {author} {\bibfnamefont {M.~P.}\ \bibnamefont {Kostylev}},\ }\href {\doibase
  10.1063/1.2834714} {\bibfield  {journal} {\bibinfo  {journal} {Applied
  Physics Letters}\ }\textbf {\bibinfo {volume} {92}},\ \bibinfo {eid} {022505}
  (\bibinfo {year} {2008})}\BibitemShut {NoStop}%
\bibitem [{\citenamefont {Chumak}\ \emph {et~al.}(2012)\citenamefont {Chumak},
  \citenamefont {Serga}, \citenamefont {Jungfleisch}, \citenamefont {Neb},
  \citenamefont {Bozhko}, \citenamefont {Tiberkevich},\ and\ \citenamefont
  {Hillebrands}}]{travellingSW}%
  \BibitemOpen
  \bibfield  {author} {\bibinfo {author} {\bibfnamefont {A.~V.}\ \bibnamefont
  {Chumak}}, \bibinfo {author} {\bibfnamefont {A.~A.}\ \bibnamefont {Serga}},
  \bibinfo {author} {\bibfnamefont {M.~B.}\ \bibnamefont {Jungfleisch}},
  \bibinfo {author} {\bibfnamefont {R.}~\bibnamefont {Neb}}, \bibinfo {author}
  {\bibfnamefont {D.~A.}\ \bibnamefont {Bozhko}}, \bibinfo {author}
  {\bibfnamefont {V.~S.}\ \bibnamefont {Tiberkevich}}, \ and\ \bibinfo {author}
  {\bibfnamefont {B.}~\bibnamefont {Hillebrands}},\ }\href {\doibase
  10.1063/1.3689787} {\bibfield  {journal} {\bibinfo  {journal} {Applied
  Physics Letters}\ }\textbf {\bibinfo {volume} {100}},\ \bibinfo {eid}
  {082405} (\bibinfo {year} {2012})}\BibitemShut {NoStop}%
\bibitem [{\citenamefont {Ando}\ and\ \citenamefont
  {Saitoh}(2011)}]{SaitohBistable}%
  \BibitemOpen
  \bibfield  {author} {\bibinfo {author} {\bibfnamefont {K.}~\bibnamefont
  {Ando}}\ and\ \bibinfo {author} {\bibfnamefont {E.}~\bibnamefont {Saitoh}},\
  }\href@noop {} {\enquote {\bibinfo {title} {{Bistable Spin Pumping Memory
  Effect}},}\ } (\bibinfo {year} {2011}),\ \Eprint
  {http://arxiv.org/abs/1112.1596v1} {arXiv:1112.1596v1} \BibitemShut {NoStop}%
\bibitem [{\citenamefont {Kurebayashi}\ \emph
  {et~al.}(2011{\natexlab{a}})\citenamefont {Kurebayashi}, \citenamefont
  {Dzyapko}, \citenamefont {Demidov}, \citenamefont {Fang}, \citenamefont
  {Ferguson},\ and\ \citenamefont {Demokritov}}]{Kurebayashi2011nmat}%
  \BibitemOpen
  \bibfield  {author} {\bibinfo {author} {\bibfnamefont {H.}~\bibnamefont
  {Kurebayashi}}, \bibinfo {author} {\bibfnamefont {O.}~\bibnamefont
  {Dzyapko}}, \bibinfo {author} {\bibfnamefont {V.~E.}\ \bibnamefont
  {Demidov}}, \bibinfo {author} {\bibfnamefont {D.}~\bibnamefont {Fang}},
  \bibinfo {author} {\bibfnamefont {A.~J.}\ \bibnamefont {Ferguson}}, \ and\
  \bibinfo {author} {\bibfnamefont {S.~O.}\ \bibnamefont {Demokritov}},\
  }\href@noop {} {\bibfield  {journal} {\bibinfo  {journal} {Nat. Mater.}\
  }\textbf {\bibinfo {volume} {10}},\ \bibinfo {pages} {660} (\bibinfo {year}
  {2011}{\natexlab{a}})}\BibitemShut {NoStop}%
\bibitem [{\citenamefont {Sandweg}\ \emph {et~al.}(2011)\citenamefont
  {Sandweg}, \citenamefont {Kajiwara}, \citenamefont {Chumak}, \citenamefont
  {Serga}, \citenamefont {Vasyuchka}, \citenamefont {Jungfleisch},
  \citenamefont {Saitoh},\ and\ \citenamefont
  {Hillebrands}}]{HillebrandsSPmagnons}%
  \BibitemOpen
  \bibfield  {author} {\bibinfo {author} {\bibfnamefont {C.~W.}\ \bibnamefont
  {Sandweg}}, \bibinfo {author} {\bibfnamefont {Y.}~\bibnamefont {Kajiwara}},
  \bibinfo {author} {\bibfnamefont {A.~V.}\ \bibnamefont {Chumak}}, \bibinfo
  {author} {\bibfnamefont {A.~A.}\ \bibnamefont {Serga}}, \bibinfo {author}
  {\bibfnamefont {V.~I.}\ \bibnamefont {Vasyuchka}}, \bibinfo {author}
  {\bibfnamefont {M.~B.}\ \bibnamefont {Jungfleisch}}, \bibinfo {author}
  {\bibfnamefont {E.}~\bibnamefont {Saitoh}}, \ and\ \bibinfo {author}
  {\bibfnamefont {B.}~\bibnamefont {Hillebrands}},\ }\href {\doibase
  10.1103/PhysRevLett.106.216601} {\bibfield  {journal} {\bibinfo  {journal}
  {Phys. Rev. Lett.}\ }\textbf {\bibinfo {volume} {106}},\ \bibinfo {pages}
  {216601} (\bibinfo {year} {2011})}\BibitemShut {NoStop}%
\bibitem [{\citenamefont {Kurebayashi}\ \emph
  {et~al.}(2011{\natexlab{b}})\citenamefont {Kurebayashi}, \citenamefont
  {Dzyapko}, \citenamefont {Demidov}, \citenamefont {Fang}, \citenamefont
  {Ferguson},\ and\ \citenamefont {Demokritov}}]{DemoParametric}%
  \BibitemOpen
  \bibfield  {author} {\bibinfo {author} {\bibfnamefont {H.}~\bibnamefont
  {Kurebayashi}}, \bibinfo {author} {\bibfnamefont {O.}~\bibnamefont
  {Dzyapko}}, \bibinfo {author} {\bibfnamefont {V.~E.}\ \bibnamefont
  {Demidov}}, \bibinfo {author} {\bibfnamefont {D.}~\bibnamefont {Fang}},
  \bibinfo {author} {\bibfnamefont {A.~J.}\ \bibnamefont {Ferguson}}, \ and\
  \bibinfo {author} {\bibfnamefont {S.~O.}\ \bibnamefont {Demokritov}},\ }\href
  {\doibase 10.1063/1.3652911} {\bibfield  {journal} {\bibinfo  {journal}
  {Applied Physics Letters}\ }\textbf {\bibinfo {volume} {99}},\ \bibinfo {eid}
  {162502} (\bibinfo {year} {2011}{\natexlab{b}})}\BibitemShut {NoStop}%
\bibitem [{\citenamefont {Jungfleisch}\ \emph {et~al.}(2011)\citenamefont
  {Jungfleisch}, \citenamefont {Chumak}, \citenamefont {Vasyuchka},
  \citenamefont {Serga}, \citenamefont {Obry}, \citenamefont {Schultheiss},
  \citenamefont {Beck}, \citenamefont {Karenowska}, \citenamefont {Saitoh},\
  and\ \citenamefont {Hillebrands}}]{2mag}%
  \BibitemOpen
  \bibfield  {author} {\bibinfo {author} {\bibfnamefont {M.~B.}\ \bibnamefont
  {Jungfleisch}}, \bibinfo {author} {\bibfnamefont {A.~V.}\ \bibnamefont
  {Chumak}}, \bibinfo {author} {\bibfnamefont {V.~I.}\ \bibnamefont
  {Vasyuchka}}, \bibinfo {author} {\bibfnamefont {A.~A.}\ \bibnamefont
  {Serga}}, \bibinfo {author} {\bibfnamefont {B.}~\bibnamefont {Obry}},
  \bibinfo {author} {\bibfnamefont {H.}~\bibnamefont {Schultheiss}}, \bibinfo
  {author} {\bibfnamefont {P.~A.}\ \bibnamefont {Beck}}, \bibinfo {author}
  {\bibfnamefont {A.~D.}\ \bibnamefont {Karenowska}}, \bibinfo {author}
  {\bibfnamefont {E.}~\bibnamefont {Saitoh}}, \ and\ \bibinfo {author}
  {\bibfnamefont {B.}~\bibnamefont {Hillebrands}},\ }\href {\doibase
  10.1063/1.3658398} {\bibfield  {journal} {\bibinfo  {journal} {Applied
  Physics Letters}\ }\textbf {\bibinfo {volume} {99}},\ \bibinfo {eid} {182512}
  (\bibinfo {year} {2011})}\BibitemShut {NoStop}%
\bibitem [{\citenamefont {Prabhakar}\ and\ \citenamefont
  {Stancil}(1999)}]{prabhakar4859}%
  \BibitemOpen
  \bibfield  {author} {\bibinfo {author} {\bibfnamefont {A.}~\bibnamefont
  {Prabhakar}}\ and\ \bibinfo {author} {\bibfnamefont {D.~D.}\ \bibnamefont
  {Stancil}},\ }\href {\doibase 10.1063/1.370045} {\bibfield  {journal}
  {\bibinfo  {journal} {Journal of Applied Physics}\ }\textbf {\bibinfo
  {volume} {85}},\ \bibinfo {pages} {4859} (\bibinfo {year}
  {1999})}\BibitemShut {NoStop}%
\bibitem [{Note1()}]{Note1}%
  \BibitemOpen
  \bibinfo {note} {The characteristic impedance of the microstrip is designed
  with respect to the source impedance. By taking into account the geometric
  dimensions of the line, the electrical properties of the line (Au), and the
  permittivity of the substrate (alumina) we have realized a microstrip line
  with an impedance of 50 Ohm. In order to create a maximum current through the
  microstrip and therefore a maximum coupling (in the frequency range that we
  have used), the end of the transmission line has been shorted.}\BibitemShut
  {Stop}%
\bibitem [{\citenamefont {Burrowes}\ \emph {et~al.}(2012)\citenamefont
  {Burrowes}, \citenamefont {Heinrich}, \citenamefont {Kardasz}, \citenamefont
  {Montoya}, \citenamefont {Girt}, \citenamefont {Sun}, \citenamefont {Song},\
  and\ \citenamefont {Wu}}]{PLD_YIG}%
  \BibitemOpen
  \bibfield  {author} {\bibinfo {author} {\bibfnamefont {C.}~\bibnamefont
  {Burrowes}}, \bibinfo {author} {\bibfnamefont {B.}~\bibnamefont {Heinrich}},
  \bibinfo {author} {\bibfnamefont {B.}~\bibnamefont {Kardasz}}, \bibinfo
  {author} {\bibfnamefont {E.~A.}\ \bibnamefont {Montoya}}, \bibinfo {author}
  {\bibfnamefont {E.}~\bibnamefont {Girt}}, \bibinfo {author} {\bibfnamefont
  {Y.}~\bibnamefont {Sun}}, \bibinfo {author} {\bibfnamefont {Y.-Y.}\
  \bibnamefont {Song}}, \ and\ \bibinfo {author} {\bibfnamefont
  {M.}~\bibnamefont {Wu}},\ }\href {\doibase 10.1063/1.3690918} {\bibfield
  {journal} {\bibinfo  {journal} {Applied Physics Letters}\ }\textbf {\bibinfo
  {volume} {100}},\ \bibinfo {eid} {092403} (\bibinfo {year}
  {2012})}\BibitemShut {NoStop}%
\bibitem [{\citenamefont {Wang}\ \emph {et~al.}(2011)\citenamefont {Wang},
  \citenamefont {Sun}, \citenamefont {Song}, \citenamefont {Wu}, \citenamefont
  {Schulthei\ss}, \citenamefont {Pearson},\ and\ \citenamefont
  {Hoffmann}}]{WangAPL2011}%
  \BibitemOpen
  \bibfield  {author} {\bibinfo {author} {\bibfnamefont {Z.}~\bibnamefont
  {Wang}}, \bibinfo {author} {\bibfnamefont {Y.}~\bibnamefont {Sun}}, \bibinfo
  {author} {\bibfnamefont {Y.-Y.}\ \bibnamefont {Song}}, \bibinfo {author}
  {\bibfnamefont {M.}~\bibnamefont {Wu}}, \bibinfo {author} {\bibfnamefont
  {H.}~\bibnamefont {Schulthei\ss}}, \bibinfo {author} {\bibfnamefont {J.~E.}\
  \bibnamefont {Pearson}}, \ and\ \bibinfo {author} {\bibfnamefont
  {A.}~\bibnamefont {Hoffmann}},\ }\href {\doibase 10.1063/1.3654148}
  {\bibfield  {journal} {\bibinfo  {journal} {Applied Physics Letters}\
  }\textbf {\bibinfo {volume} {99}},\ \bibinfo {eid} {162511} (\bibinfo {year}
  {2011})}\BibitemShut {NoStop}%
\bibitem [{\citenamefont {Ando}\ \emph {et~al.}(2010)\citenamefont {Ando},
  \citenamefont {Kajiwara}, \citenamefont {Sasage}, \citenamefont {Uchida},\
  and\ \citenamefont {Saitoh}}]{AndoIEEE2010}%
  \BibitemOpen
  \bibfield  {author} {\bibinfo {author} {\bibfnamefont {K.}~\bibnamefont
  {Ando}}, \bibinfo {author} {\bibfnamefont {Y.}~\bibnamefont {Kajiwara}},
  \bibinfo {author} {\bibfnamefont {K.}~\bibnamefont {Sasage}}, \bibinfo
  {author} {\bibfnamefont {K.}~\bibnamefont {Uchida}}, \ and\ \bibinfo {author}
  {\bibfnamefont {E.}~\bibnamefont {Saitoh}},\ }\href@noop {} {\bibfield
  {journal} {\bibinfo  {journal} {IEEE Transactions on Magnetics}\ }\textbf
  {\bibinfo {volume} {46}},\ \bibinfo {pages} {3694 } (\bibinfo {year}
  {2010})}\BibitemShut {NoStop}%
\bibitem [{\citenamefont {Padr\'{o}n-Hern\'{a}ndez}\ \emph
  {et~al.}(2011)\citenamefont {Padr\'{o}n-Hern\'{a}ndez}, \citenamefont
  {Azevedo},\ and\ \citenamefont {Rezende}}]{AzevedoAPL2011}%
  \BibitemOpen
  \bibfield  {author} {\bibinfo {author} {\bibfnamefont {E.}~\bibnamefont
  {Padr\'{o}n-Hern\'{a}ndez}}, \bibinfo {author} {\bibfnamefont
  {A.}~\bibnamefont {Azevedo}}, \ and\ \bibinfo {author} {\bibfnamefont
  {S.~M.}\ \bibnamefont {Rezende}},\ }\href {\doibase 10.1063/1.3660586}
  {\bibfield  {journal} {\bibinfo  {journal} {Applied Physics Letters}\
  }\textbf {\bibinfo {volume} {99}},\ \bibinfo {eid} {192511} (\bibinfo {year}
  {2011})}\BibitemShut {NoStop}%
\bibitem [{\citenamefont {Padr\'on-Hern\'andez}\ \emph
  {et~al.}(2011)\citenamefont {Padr\'on-Hern\'andez}, \citenamefont {Azevedo},\
  and\ \citenamefont {Rezende}}]{AzevedoPRL2011}%
  \BibitemOpen
  \bibfield  {author} {\bibinfo {author} {\bibfnamefont {E.}~\bibnamefont
  {Padr\'on-Hern\'andez}}, \bibinfo {author} {\bibfnamefont {A.}~\bibnamefont
  {Azevedo}}, \ and\ \bibinfo {author} {\bibfnamefont {S.~M.}\ \bibnamefont
  {Rezende}},\ }\href {\doibase 10.1103/PhysRevLett.107.197203} {\bibfield
  {journal} {\bibinfo  {journal} {Phys. Rev. Lett.}\ }\textbf {\bibinfo
  {volume} {107}},\ \bibinfo {pages} {197203} (\bibinfo {year}
  {2011})}\BibitemShut {NoStop}%
\bibitem [{\citenamefont {Youssef}\ \emph {et~al.}(2010)\citenamefont
  {Youssef}, \citenamefont {Castel}, \citenamefont {Vukadinovic},\ and\
  \citenamefont {Labrune}}]{jbyJAP2010}%
  \BibitemOpen
  \bibfield  {author} {\bibinfo {author} {\bibfnamefont {J.~B.}\ \bibnamefont
  {Youssef}}, \bibinfo {author} {\bibfnamefont {V.}~\bibnamefont {Castel}},
  \bibinfo {author} {\bibfnamefont {N.}~\bibnamefont {Vukadinovic}}, \ and\
  \bibinfo {author} {\bibfnamefont {M.}~\bibnamefont {Labrune}},\ }\href
  {\doibase 10.1063/1.3475646} {\bibfield  {journal} {\bibinfo  {journal}
  {Journal of Applied Physics}\ }\textbf {\bibinfo {volume} {108}},\ \bibinfo
  {eid} {063909} (\bibinfo {year} {2010})}\BibitemShut {NoStop}%
\bibitem [{\citenamefont {Stancil}\ and\ \citenamefont
  {Prabhakar}(2009)}]{SWstancil}%
  \BibitemOpen
  \bibfield  {author} {\bibinfo {author} {\bibfnamefont {D.~D.}\ \bibnamefont
  {Stancil}}\ and\ \bibinfo {author} {\bibfnamefont {A.}~\bibnamefont
  {Prabhakar}},\ }\href@noop {} {\emph {\bibinfo {title} {Spin Waves: Theory
  and Applications}}}\ (\bibinfo  {publisher} {Springer},\ \bibinfo {address}
  {New York},\ \bibinfo {year} {2009})\BibitemShut {NoStop}%
\bibitem [{\citenamefont {Sandweg}\ \emph {et~al.}(2010)\citenamefont
  {Sandweg}, \citenamefont {Kajiwara}, \citenamefont {Ando}, \citenamefont
  {Saitoh},\ and\ \citenamefont {Hillebrands}}]{HilleSelection}%
  \BibitemOpen
  \bibfield  {author} {\bibinfo {author} {\bibfnamefont {C.~W.}\ \bibnamefont
  {Sandweg}}, \bibinfo {author} {\bibfnamefont {Y.}~\bibnamefont {Kajiwara}},
  \bibinfo {author} {\bibfnamefont {K.}~\bibnamefont {Ando}}, \bibinfo {author}
  {\bibfnamefont {E.}~\bibnamefont {Saitoh}}, \ and\ \bibinfo {author}
  {\bibfnamefont {B.}~\bibnamefont {Hillebrands}},\ }\href {\doibase
  10.1063/1.3528207} {\bibfield  {journal} {\bibinfo  {journal} {Applied
  Physics Letters}\ }\textbf {\bibinfo {volume} {97}},\ \bibinfo {eid} {252504}
  (\bibinfo {year} {2010})}\BibitemShut {NoStop}%
\bibitem [{\citenamefont {Suhl}(1957)}]{Suhl}%
  \BibitemOpen
  \bibfield  {author} {\bibinfo {author} {\bibfnamefont {H.}~\bibnamefont
  {Suhl}},\ }\href {\doibase 10.1016/0022-3697(57)90010-0} {\bibfield
  {journal} {\bibinfo  {journal} {Journal of Physics and Chemistry of Solids}\
  }\textbf {\bibinfo {volume} {1}},\ \bibinfo {pages} {209 } (\bibinfo {year}
  {1957})}\BibitemShut {NoStop}%
\bibitem [{\citenamefont {Weiss}(1958)}]{Weiss}%
  \BibitemOpen
  \bibfield  {author} {\bibinfo {author} {\bibfnamefont {M.~T.}\ \bibnamefont
  {Weiss}},\ }\href {\doibase 10.1103/PhysRevLett.1.239} {\bibfield  {journal}
  {\bibinfo  {journal} {Phys. Rev. Lett.}\ }\textbf {\bibinfo {volume} {1}},\
  \bibinfo {pages} {239} (\bibinfo {year} {1958})}\BibitemShut {NoStop}%
\bibitem [{\citenamefont {Guan}\ \emph {et~al.}(2007)\citenamefont {Guan},
  \citenamefont {Bailey}, \citenamefont {Vescovo}, \citenamefont {Kao},\ and\
  \citenamefont {Arena}}]{thetaAngle}%
  \BibitemOpen
  \bibfield  {author} {\bibinfo {author} {\bibfnamefont {Y.}~\bibnamefont
  {Guan}}, \bibinfo {author} {\bibfnamefont {W.~E.}\ \bibnamefont {Bailey}},
  \bibinfo {author} {\bibfnamefont {E.}~\bibnamefont {Vescovo}}, \bibinfo
  {author} {\bibfnamefont {C.-C.}\ \bibnamefont {Kao}}, \ and\ \bibinfo
  {author} {\bibfnamefont {D.~A.}\ \bibnamefont {Arena}},\ }\href {\doibase
  10.1016/j.jmmm.2006.10.1111} {\bibfield  {journal} {\bibinfo  {journal}
  {Journal of Magnetism and Magnetic Materials}\ }\textbf {\bibinfo {volume}
  {312}},\ \bibinfo {pages} {374} (\bibinfo {year} {2007})}\BibitemShut
  {NoStop}%
\bibitem [{\citenamefont {Kittel}(1948)}]{Kittel}%
  \BibitemOpen
  \bibfield  {author} {\bibinfo {author} {\bibfnamefont {C.}~\bibnamefont
  {Kittel}},\ }\href {\doibase 10.1103/PhysRev.73.155} {\bibfield  {journal}
  {\bibinfo  {journal} {Phys. Rev.}\ }\textbf {\bibinfo {volume} {73}},\
  \bibinfo {pages} {155} (\bibinfo {year} {1948})}\BibitemShut {NoStop}%
\bibitem [{\citenamefont {Kalinikos}\ \emph {et~al.}(1990)\citenamefont
  {Kalinikos}, \citenamefont {Kostylev}, \citenamefont {Kozhus},\ and\
  \citenamefont {Slavin}}]{SWdispExchange}%
  \BibitemOpen
  \bibfield  {author} {\bibinfo {author} {\bibfnamefont {B.~A.}\ \bibnamefont
  {Kalinikos}}, \bibinfo {author} {\bibfnamefont {M.~P.}\ \bibnamefont
  {Kostylev}}, \bibinfo {author} {\bibfnamefont {N.~V.}\ \bibnamefont
  {Kozhus}}, \ and\ \bibinfo {author} {\bibfnamefont {A.~N.}\ \bibnamefont
  {Slavin}},\ }\href {http://stacks.iop.org/0953-8984/2/i=49/a=012} {\bibfield
  {journal} {\bibinfo  {journal} {Journal of Physics: Condensed Matter}\
  }\textbf {\bibinfo {volume} {2}},\ \bibinfo {pages} {9861} (\bibinfo {year}
  {1990})}\BibitemShut {NoStop}%
\bibitem [{\citenamefont {Ando}\ \emph {et~al.}(2009)\citenamefont {Ando},
  \citenamefont {Yoshino},\ and\ \citenamefont {Saitoh}}]{andoOptimum}%
  \BibitemOpen
  \bibfield  {author} {\bibinfo {author} {\bibfnamefont {K.}~\bibnamefont
  {Ando}}, \bibinfo {author} {\bibfnamefont {T.}~\bibnamefont {Yoshino}}, \
  and\ \bibinfo {author} {\bibfnamefont {E.}~\bibnamefont {Saitoh}},\ }\href
  {\doibase 10.1063/1.3119314} {\bibfield  {journal} {\bibinfo  {journal}
  {Applied Physics Letters}\ }\textbf {\bibinfo {volume} {94}},\ \bibinfo {eid}
  {152509} (\bibinfo {year} {2009})}\BibitemShut {NoStop}%
\bibitem [{\citenamefont {Boardman}\ and\ \citenamefont
  {Nikitov}(1988)}]{PowerThresNikitov1988}%
  \BibitemOpen
  \bibfield  {author} {\bibinfo {author} {\bibfnamefont {A.~D.}\ \bibnamefont
  {Boardman}}\ and\ \bibinfo {author} {\bibfnamefont {S.~A.}\ \bibnamefont
  {Nikitov}},\ }\href {\doibase 10.1103/PhysRevB.38.11444} {\bibfield
  {journal} {\bibinfo  {journal} {Phys. Rev. B}\ }\textbf {\bibinfo {volume}
  {38}},\ \bibinfo {pages} {11444} (\bibinfo {year} {1988})}\BibitemShut
  {NoStop}%
\bibitem [{\citenamefont {Jun}\ \emph {et~al.}(1997)\citenamefont {Jun},
  \citenamefont {Nikitov}, \citenamefont {Marcelli},\ and\ \citenamefont
  {Gasperis}}]{TMSNikitov}%
  \BibitemOpen
  \bibfield  {author} {\bibinfo {author} {\bibfnamefont {S.}~\bibnamefont
  {Jun}}, \bibinfo {author} {\bibfnamefont {S.~A.}\ \bibnamefont {Nikitov}},
  \bibinfo {author} {\bibfnamefont {R.}~\bibnamefont {Marcelli}}, \ and\
  \bibinfo {author} {\bibfnamefont {P.~D.}\ \bibnamefont {Gasperis}},\ }\href
  {\doibase 10.1063/1.363869} {\bibfield  {journal} {\bibinfo  {journal}
  {Journal of Applied Physics}\ }\textbf {\bibinfo {volume} {81}},\ \bibinfo
  {pages} {1341} (\bibinfo {year} {1997})}\BibitemShut {NoStop}%
\end{thebibliography}%

\end{document}